  \documentstyle[aps,12pt,tighten,epsf]{revtex}

\begin{document}
\newcommand{\bS}     {\mbox{\boldmath $S$}}
\newcommand{\bT}     {\mbox{\boldmath $T$}}
\newcommand{\bR}     {\mbox{\boldmath $R$}}
\newcommand{\bU}     {\mbox{\boldmath $U$}}
\newcommand{\bsigma} {\mbox{\boldmath $\sigma$}}
\newcommand{\bk}     {\mbox{\boldmath $k$}}
\newcommand{\one}    {\mbox{$\openone$}}
\newcommand{\br}     {\mbox{\boldmath $r$}}
\newcommand{\bt}     {\mbox{\boldmath $t$}}
\newcommand{\half}   {\mbox{${\textstyle \frac{1}{2}}$}}


\title{Decomposition of Resonant Scatterers by Surfaces of Section}

\author{Ra\'ul O. Vallejos$^1$ and Alfredo M. Ozorio de Almeida$^2$}

\address{{$^1$}Instituto de F\'{\i}sica, 
   Universidade do Estado do Rio de Janeiro, \\
   Rua S\~ao Francisco Xavier 524, CEP 20559-900 Rio de Janeiro, 
   Brazil \\
   {\rm e-mail: raul@dfnae1.fis.uerj.br} \\
   $^2$Centro Brasileiro de Pesquisas F\'{\i}sicas, \\
   Rua Xavier Sigaud 150, CEP 22290-180 Rio de Janeiro, Brazil \\
   {\rm e-mail: ozorio@cbpf.br} }

\date{\today}

\maketitle

\begin{abstract} 
Scattering on the energy shell is viewed here as the relation between 
the bound states of the Hamiltonian, restricted to sections on leads
that are asymptotically independent, far away from the interaction 
region. The decomposition is achieved by sectioning this region and
adding new leads, thus generating two new scatterers. So a resonant
scatterer, whose $\bS$-matrix has sharp energy peaks, can be resolved 
into a pair of scatterers with smooth energy dependence. The resonant
behaviour is concentrated in a spectral determinant obtained from a
dissipative section map. The semiclassical limit of this theory coincides
with the orbit resummation previously proposed by Georgeot and Prange.
A numerical example for a semiseparable scatterer is investigated,
revealing the accurate portrayal of the Wigner time delay by the
spectral determinant.
\end{abstract}

\draft 




\section{Introduction}


Surfaces of section were conceived by Poincar\'e as a means to
decompose bounded classical motion into discrete maps of the section
onto itself.  Even if the longterm motion is chaotic in some sense,
there are many important cases where the single return map is well
behaved, so that the complex intermingling of trajectories results from
repeated iterations.  If a section is badly chosen, the first
return map may well be singular and discontinuous with fractal
boundaries between continuous regions.  Evidently, such a Poincar\'e
map will not be a useful tool for the study of the classical motion.
There are many systems for which no single surface exists that leads to
a well behaved Poincar\'e map, though these have not been much studied
for obvious reasons.

It is only with the work of Bogomolny \cite{bogomolny92} that surfaces
of section where brought into quantum mechanics.  At first this was
viewed as an essentially semiclassical extrapolation of Poincar\'e's
method, while limiting the section to position space instead of the
classical phase space.  However, the approach of Rouvinez and Smilansky
\cite{rouvinez95} and Prosen \cite{prosen95}, allows us to define the
transformation, undergone by the finite Hilbert space corresponding to
the section, as the product of two unitary
scattering matrices.  These describe the two alternative unbounded
motions, obtained by substituting either side of the system by a
semi-infinite tube.  Defining the scattering operators as $\bS_L(E)$
and $\bS_R(E)$, where $E$ is the energy of the open system, the
condition for a bound state to exist at a certain energy is just that
$\bS_L \bS_R$, or $\bS_R \bS_L$ has a unit eigenvalue.

\begin{figure}[ht]
\epsfysize=10.0cm
\epsfbox[100 165 492 629]{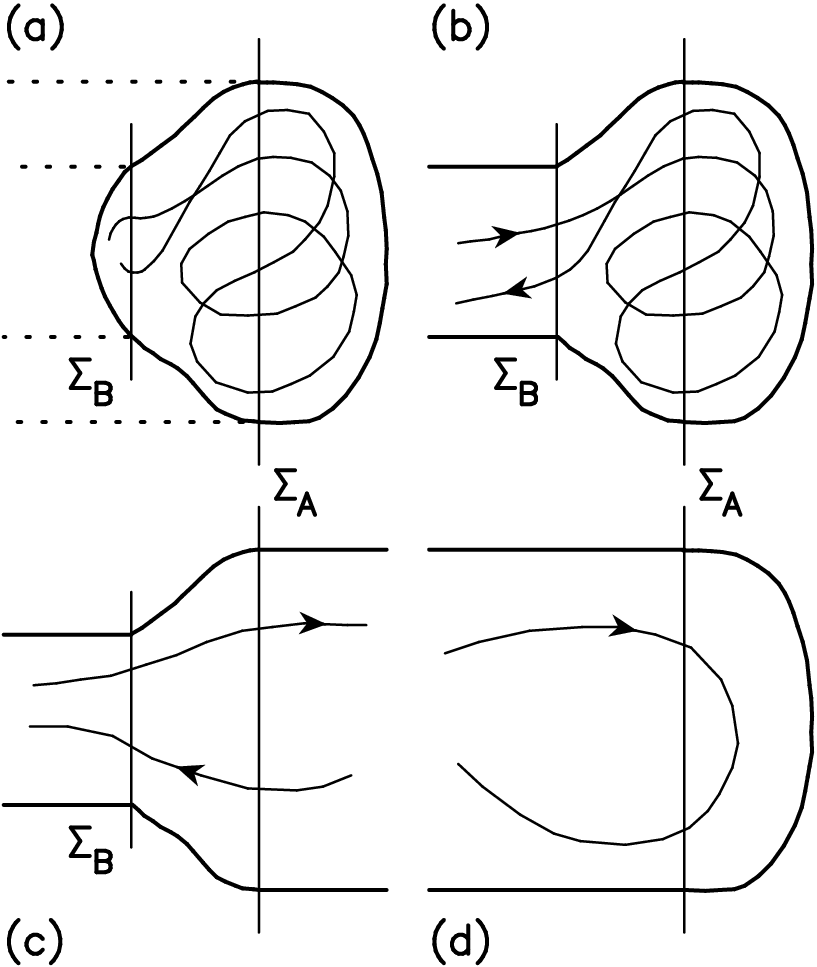}
\vspace{1cm}
\caption{A single return of the orbit to $\Sigma_B$ corresponds to
multiple iterations of the Poincar\'e map for $\Sigma_A$.  The surface
of section $\Sigma_A$ decomposes the resonant scatterer in (b) into two
nonresonant scattering systems (c) and (d).  Accordingly, the
scattering trajectories are complex in (b) but simple in both (c) and
(d). Thick curves represent a single contour of the potential energy.}
\label{fig1}
\end{figure}

The observation that a ``bad'' choice of section results in a maze of
contributing orbits has a parallel in quantum mechanics that the
correponding 
scattering exhibits multiple resonances. For example, consider the
system in Fig.~\ref{fig1}(a). For the good choice of section, the
classical orbits return to $\Sigma_A$ after a short time on the right
side, whereas the return map for $\Sigma_B$ will be very
complex. This corresponds to a smooth energy dependence of all the
elements of the scattering matrix $\bS_{LA}$ (for the open system 
obtained by joining an open tube to the left of $\Sigma_A$), 
as opposed to a spiky
energy dependence of $\bS_{LB}$ due to many resonances.
Let us now consider the case where the system under study is actually
the open system depicted in Fig.~\ref{fig1}(b), obtained by joining a 
narrow open tube to the left of $\Sigma_B$ in system (a). The
``natural section'' that describes this new scattering problem
would then be $\Sigma_B$, but having seen that this is a ``bad choice''
for the closed system, we would also like to bypass it in the description
of the new open system (b).
It may then be advantageous
to decompose this open resonant system using a ``good'' section
$\Sigma_A$, just as in the case of the closed system. Indeed, we may
consider the latter as the limit of a family of open systems in which
the ``bad'' section is pushed evermore to the left until the opening
at $\Sigma_B$ has zero width. The system (b) will
then be decomposed into the auxiliary systems of Fig.~\ref{fig1}(c) and
(d).  The novelty with respect to the closed problem is that now there
are modes entering the scattering region between $\Sigma_A$ and
$\Sigma_B$ from the tubes both to the left and to the right of 
Fig.~\ref{fig1}(c). The
important point is that the classical orbits remain only a short time
in the scattering region, leading to simple mappings among the surfaces
$\Sigma_A$ and $\Sigma_B$ and nonresonant energy dependence of the
$\bS$-matrices.

The purpose of this paper is to present the general method for the
decomposition of a resonant scatterer into two less resonant scatterers
by means of a surface of section. If the optimal decomposition renders
both the resulting scatterers nonresonant, the objective is reached and
the resonances will result from multiple scattering across the surface,
as will be explicitly shown in Sec.~\ref{sec:resonance}. 
If there is no choice of
section for which one of the components is not resonant, we can clearly
decompose it in the same manner and so on until all the components are
nonresonant. A similar sequence of decompositions may be required for a
bound system. There may not be any surface that avoids resonances in 
$\bS_L$ or $\bS_R$. The further decomposition of either scatterer proceeds
exactly as for initially open systems.

First we review multichannel scattering theory on the energy shell,
establishing the notation. In Sec.~\ref{sec:decomposition},
we decompose the scatterer through a surface of section. The way that
the resonant structure emerges from the decomposed scatterers is
described in the following section. Particular attention is given to
the time delay, as an overall indicator of resonance. Semiclassical
approximations are discussed in Sec.~\ref{sec:semiclassical}. These
are not valid across a discontinuity, so that the local treatment for
its traversal follows in Sec.~\ref{sec:discontinuities}. We conclude
with a numerical calculation for a semiseparable scatterer which allows
us to evaluate the discontinuous approximation advanced in the previous
section. We also verify that the localization of resonance peaks of the
time delay is accurately determined by the determinant of a quantum
Poincar\'e map for the open system.


\section{Multichannel scattering}


\label{sec:multichannel}

We here review the theory of scattering on the energy shell so as to
define the notation. Careful definition of the context determines
the scope of the following theory in its present form. A general
scatterer connects regions where asymptotic states, modes or channels,
propagate independently.  Without loss of generality, we may model the
scatterer as a cavity connecting entrance leads\cite{lewenkopf91}.
In many important cases, such as in mesoscopic devices, this
structure is immediately evident, but even the scattering of waves from
a localized object may be considered as radial backscattering along a
tube of angular width $2\pi$.  Defining $x_J$ as the longitudinal
coordinate along the $J$-th lead and letting $y_J$ represent the
transverse coordinates, the asymptotic region is specified by the
requirement that, as $x_J \to \infty$, the Hamiltonian
$H(x_J,y_J,p_{xJ},p_{yJ})$ becomes independent of $x_J$:
\begin{equation}
H(x_J,y_J,p_{xJ},p_{yJ}) 
\stackrel{x_J \to \infty}{\longrightarrow} 
       \frac{p_{xJ}^2}{2m} + H_J(y_J,p_{yJ}) \; .
\label{hamiltonian}
\end{equation}
Here we assume that the scattering region lies near the origin of
$x_J$ and that $x_J$ is always positive in the asymptotic region.
We also assume bounded motion in the transverse directions, implying that
$H_J(y_J,p_{yJ})$ has a discrete infinite spectrum of states
$|n_J \}$ with energies $\epsilon_{nJ}$.

The Hilbert space for the stationary solutions of the Schr\"odinger
equation are decomposed into orthogonal subspaces in the asymptotic
region of each lead. The $J$-th subspace is a superposition of states
\begin{equation}
\langle x_J,y_J | n_J \rangle ^\pm =
\frac{1}{\sqrt{k_{nJ}}} \exp(\pm i k_{nJ} x_J)\{y_J|n_J \} \equiv 
\frac{1}{\sqrt{k_{nJ}}} \{y_J|n(x_J) \}^\pm \; ,
\label{waves}
\end{equation}
where
\begin{equation}
k_{nJ} = \sqrt{\frac{2m(E-\epsilon_{nJ})}{\hbar^2}} \; .
\label{wavenumbers}
\end{equation}
In other words, the interaction between the transverse states
$| n_J \}$ is switched off as $x_J$ enters the asymptotic region.
There remains only a trivial exponential dependence of the full
state $| n_J \rangle$ on the longitudinal coordinate.
(We partially adopt the notation of Prosen in \cite{prosen95}.)
%
%

The states for which $k_{nJ}$ is real are referred to as {\em open}
channels, otherwise they are {\em closed}. For any given energy $E$,
there will be a finite number, $\Lambda_J$, of open channels.
The closed channels with $n_J > \Lambda_J$ take the form of 
{\em evanescent modes},
\begin{equation}
\langle x_J,y_J | n_J \rangle^e =
\frac{1}{\sqrt{i|k_{nJ}|}} \exp(-| k_{nJ} | x_J)\{y_J|n_J \} \; ,
\label{evanescent}
\end{equation}
for the wave to be bounded as $x_J \to \infty$. However, we shall
usually keep to (\ref{waves}) for both open and closed channels.

A general solution $|\psi \rangle$ of the stationary Schr\"odinger
equation is decomposed in the $J$'th tube in terms of transverse
amplitudes $\{n_J|\psi_J\}$
\begin{equation}
\langle x_J,y_J | \psi \rangle = \sum_{n=0}^\infty
\frac{1}{\sqrt{k_{nJ}}} 
\left[ \{y_J|n(x_J)\}^{+ \; +}\{n_J|\psi_J \} + 
       \{y_J|n(x_J)\}^{- \; -}\{n_J|\psi_J \} \right] \; .
\label{general}
\end{equation}
Here we could also have decomposed $|\psi\rangle$ with the
coefficients $^\pm\{n(x_J)|\psi\}$ to maintain
the symmetry of the bracket notation, but it is easier to
use a basis that is independent of $x_J$.
The {\em incoming wave} $|\psi(x_J)\}^-$ and the {\em outgoing wave} 
$|\psi(x_J)\}^+$ at $x_J$ are thus defined as
\begin{equation}
|\psi(x_J)\}^\pm \equiv \sum_{n=0}^\infty 
|n(x_J)\}^{\pm \; \pm}\{n_J|\psi_J\} \; ,
\label{asymptotic}
\end{equation}
for the open channels.
The generalized $\bS$ matrix is now defined by the equation
\begin{equation}
      \left( |\psi(x_1)\}^+,|\psi(x_2)\}^+,\ldots \right) =
 \bS  \left( |\psi(x_1)\}^-,|\psi(x_2)\}^-,\ldots \right)   \; .
\label{scatdef}
\end{equation}
Conservation of current among all the leads determines that the
restriction of $\bS$ to the finite block that includes all the open
channels is a unitary matrix;
time reversal symmetry implies that the full $\bS$ matrix must be
symmetric, {\em i.e.}, $\bS_{ij}$=$\bS_{ji}$\cite{rouvinez95}.
Evidently, $\bS$ depends on the choice of $x_J$, but only up to a
translation within the asymptotic region, implemented by a diagonal 
matrix in the $|n(x_J)\}$
representation. The nontrivial energy dependence of $\bS$ will be our
main concern.


\section{Decomposition}


\label{sec:decomposition}

\begin{figure}[ht]
\epsfxsize=15cm
\epsfbox[94 323 521 467]{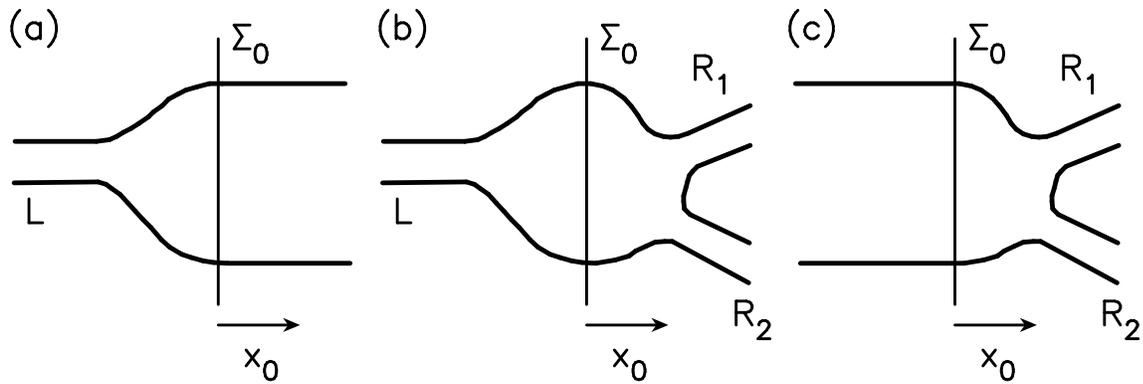}
\vspace{1cm}
\caption{The multilead scattering system.  The section $\Sigma_0$
decomposes the full scatterer (b) into the nonresonant systems (a) and
(c). Thick curves represent a single contour of the potential energy}
\label{fig2}
\end{figure}

Let us consider an arbitrary scatterer, with $J$ ($\ge 1$) leads such as
sketched in Fig.~\ref{fig2}(b). 
Cutting the scatterer by an arbitrary 
plane $\Sigma_0$, so as to avoid the leads, divides these into $R$-leads to
the right of $\Sigma_0$ and $L$-leads to the left (either of these sets
may be empty). Likewise, we divide all the channels (open and closed)
into $R$-channels and $L$-channels. Henceforth it will be immaterial
how the independent $R$-channels ($L$-channels) are subdivided among the
various $R$-leads ($L$-leads).

Define the coordinate $x_0$ growing in the direction normal to
$\Sigma_0$ and positive on the $R$-side and define the Hamiltonian
$H_0=H(x_0=0)$. Substituting $H$ by $H_0$ for all points to the right of
$\Sigma_0$, we generate a semi-infinite lead. This can be joined
onto $H$, to the left of $\Sigma_0$ so as to form the $L0$
scatterer sketched in Fig~\ref{fig2}(a). This procedure is reflected across
$\Sigma_0$ to form the $0R$ scatterer, as shown in Fig.~\ref{fig2}(c). 
The $L0$ scatterer has only one $0$-lead on the right and the states 
$|n_0\}$ are the eigenstates of $H_0 -p_{x_0}^2/2m$. These coincide with the
states to the left of the $0R$ scatterer.

We now derive the relation of the $L0$ and the $0R$ scattering to the
original problem, {\em i.e.}, the $LR$ scattering decomposed by $\Sigma_0$. 
The constant profile of the $H_0$ allows us to choose
$x_0 \to 0$ in both the  auxiliary scattering problems, which shall
be assumed henceforth.
The first step is to divide the scattering matrix for each of the
scatterers into reflection and transmision blocks, generated by the 
subdivision of the the channels of each scatterer by $\Sigma_0$:
\begin{equation}
\bS =
\left( \matrix{ \bR_{LL}   & \bT_{LR} \cr \bT_{RL} & \bR_{RR}   } \right) ,~
\bS_L=
\left( \matrix{ \bR_{LL}^0 & \bT_{L0} \cr \bT_{0L} & \bR_{00}^L } \right) ,~
\bS_R=
\left( \matrix{ \bR_{RR}^0 & \bT_{R0} \cr \bT_{0R} & \bR_{00}^R } \right) \;.
\label{matdef}
\end{equation}
To avoid confusion, we define the states related by each of these 
matrices
\begin{eqnarray}
       \left( |\psi_L(x_L)\}^+ , |\psi_{R }(x_R)\}^+\right) & = &
 \bS   \left( |\psi_L(x_L)\}^- , |\psi_{R }(x_R)\}^-\right)        \; , 
 \label{confusion1} \\
       \left( |\psi_L(x_L)\}^+ , |\psi_{L0}(x_0)\}^+\right) & = &
 \bS_L \left( |\psi_L(x_L)\}^- , |\psi_{L0}(x_0)\}^-\right)        \; , 
 \label{confusion2} \\
       \left( |\psi_R(x_R)\}^+ , |\psi_{0R}(x_0)\}^+\right) & = &
 \bS_R \left( |\psi_R(x_R)\}^- , |\psi_{0R}(x_0)\}^-\right)        \; ,
 \label{confusion3}
\end{eqnarray}
where $R$, $L$, $0R$, and $L0$ should be taken as specific instances
of the index $J$ in the definition (\ref{asymptotic}) and in preceeding 
formulae. The matrices $\bS$, $\bS_L$ and $\bS_R$ account for the full
scattering of three different systems. Our immediate task is to derive
the elements of $\bS$ from those of $\bS_L$ and $\bS_R$.

We have already identified the states on the right of the scatterer
in Fig.~\ref{fig2}(a) with those on the left of Fig.~\ref{fig2}(c), 
as well as those on the left of 
Fig.~\ref{fig2}(a) and (b). Now we match smoothly the wavefunctions of 
Fig.~\ref{fig2}(a) and (c) at $x_0=0$.
Defining the operators
\begin{equation}
\bk_J^\alpha = \sum_n |n\} k_{nJ}^\alpha \{n| \; ,
\label{kalpha}
\end{equation}
for any arbitrary power $\alpha$, we have the matching conditions 
\begin{equation}
\bk_0^{-1/2} | \psi_{L0}(x_0) \}^+  + 
\bk_0^{-1/2} | \psi_{L0}(x_0) \}^-  =
\bk_0^{-1/2} | \psi_{0R}(x_0) \}^+  +
\bk_0^{-1/2} | \psi_{0R}(x_0) \}^-  
\label{matchk1}
\end{equation}
and
\begin{equation}
 i \bk_0^{1/2} | \psi_{L0}(x_0) \}^+  - 
 i \bk_0^{1/2} | \psi_{L0}(x_0) \}^-  =
-i \bk_0^{1/2} | \psi_{0R}(x_0) \}^+  +
 i \bk_0^{1/2} | \psi_{0R}(x_0) \}^-     \; ,
\label{matchk2}
\end{equation}
which determine uniquely
\begin{equation}
 |\psi_{0R}(x_0) \}^+  = |\psi_{L0}(x_0) \}^-   
\end{equation}
and
\begin{equation}
| \psi_{L0}(x_0) \}^+  = | \psi_{0R}(x_0) \}^-  \; . 
\end{equation}
Inserting these equalities into equations (\ref{confusion2}) and 
(\ref{confusion3}), we obtain
\begin{eqnarray}
\bT_{0L}   | \psi_{L }(x_L) \}^-  +
\bR_{00}^L | \psi_{L0}(x_0) \}^-  & = & | \psi_{0R}(x_0) \}^-  \; , \\
\bT_{0R}   | \psi_{R }(x_R) \}^-  +
\bR_{00}^R | \psi_{0R}(x_0) \}^-  & = & | \psi_{L0}(x_0) \}^-  \; .
\end{eqnarray}
This system of equations for 
$| \psi_{0R}(x_0) \}^-$ and $| \psi_{L0}(x_0) \}^-$ is easily solved:
\begin{eqnarray}
| \psi_{L0}(x_0) \}^-  & = & \left[ \one-\bR_{00}^R \bR_{00}^L \right]^{-1}
                             \left[ \bR_{00}^R \bT_{0L} |\psi_L(x_L)\}^-
                                           + \bT_{0R} |\psi_R(x_R)\}^-
                             \right] \; , \\
| \psi_{0R}(x_0) \}^-  & = & \left[ \one-\bR_{00}^L \bR_{00}^R \right]^{-1}
                             \left[ \bR_{00}^L \bT_{0R} |\psi_R(x_R)\}^-
                                           + \bT_{0L} |\psi_L(x_L)\}^-
                             \right] \; ,
\end{eqnarray}
where $\one$ is the unit matrix.
Combining this result with (\ref{confusion2}) and (\ref{confusion3}) 
again, leads to
\begin{eqnarray}
& & \bR_{LL}^0 | \psi_{L}(x_L) \}^- + \nonumber \\
& & \bT_{L0} \left[ \one-\bR_{00}^R \bR_{00}^L \right]^{-1}
       \left[   \bR_{00}^R \bT_{0L} |\psi_L(x_L)\}^-
     + \bT_{0R} |\psi_R(x_R)\}^- \right] = | \psi_L (x_L) \}^+ \; , \\
& & \bR_{RR}^0 | \psi_{R}(x_R) \}^-  + \nonumber \\
& & \bT_{R0} \left[ \one-\bR_{00}^L \bR_{00}^R \right]^{-1}
       \left[   \bR_{00}^L \bT_{0R} |\psi_R(x_R)\}^-
     + \bT_{0L} |\psi_L(x_L)\}^- \right] = | \psi_R (x_R) \}^+ \; .
\end{eqnarray}
But this is just the required linear relation between 
the incoming waves $| \psi_L (x_L) \}^-$, $| \psi_R (x_R) \}^-$  and 
the outgoing waves $| \psi_L (x_L) \}^+$, $| \psi_R (x_R) \}^+$, so that
finally we have 
\begin{eqnarray}
\bR_{LL} & = & \bR_{LL}^0 + \bT_{L0} 
\left[ \one-\bR_{00}^R \bR_{00}^L \right]^{-1} \bR_{00}^R \bT_{0L}   \; ,
\label{transref1} \\
\bR_{RR} & = & \bR_{RR}^0 + \bT_{R0} 
\left[ \one-\bR_{00}^L \bR_{00}^R \right]^{-1} \bR_{00}^L \bT_{0R}   \; , 
\label{transref2} \\
\bT_{LR} & = &              \bT_{L0} 
\left[ \one-\bR_{00}^R \bR_{00}^L \right]^{-1}            \bT_{0R}   \; , 
\label{transref3} \\
\bT_{RL} & = &              \bT_{R0} 
\left[ \one-\bR_{00}^L \bR_{00}^R \right]^{-1}            \bT_{0L}   \; .
\label{transref4}  
\end{eqnarray}
\noindent
In this way, the elements of the $\bS$ matrix for the general
scatterer are completely determined by those of the scatterers
decomposed by $\Sigma_0$.

\begin{figure}[ht]
\epsfysize=5cm
\epsfbox[94 333 512 456]{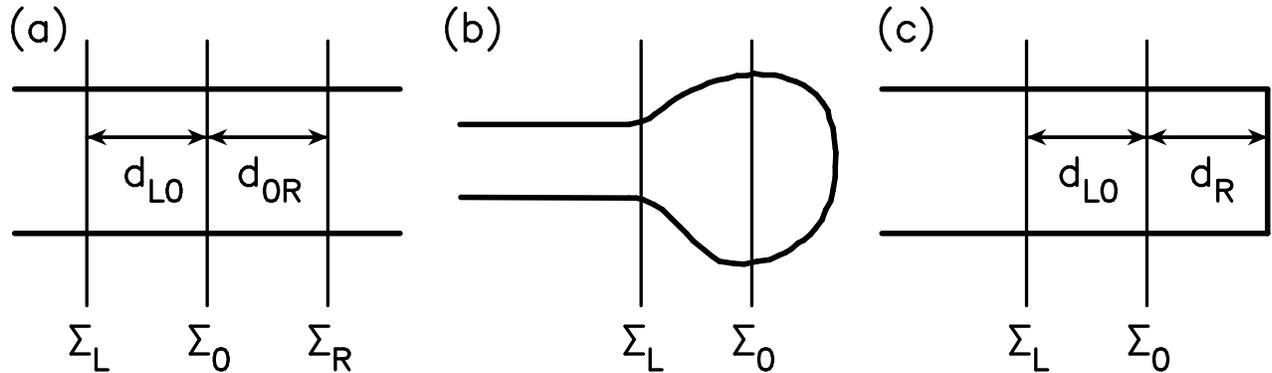}
\vspace{1cm}
\caption{Three scattering systems: a straight open tube (a), a cavity
connected to a single lead (b), and a tube closed by a hard wall (c). }
\label{fig3}\end{figure}

We now discuss some elementary examples to illustrate the foregoing
theory.
The most trivial example of a scatterer is an arbitrary section of an 
open tube. Fixing the surfaces $\Sigma_R$ and $\Sigma_L$ as in 
Fig.~\ref{fig3}(a), we can ``decompose'' it with a surface $\Sigma_0$ between
these. The matrix $\bR_{00}=0$, whereas the transmission coefficients
are diagonal: 
\begin{eqnarray}
\left( \bT_{L0} \right)_{nn}   & = & \exp(ik_n d_{L0})              =   
\left( \bT_{0L} \right)_{nn}^*                           \; , 
\label{diagonal1} \\
\left( \bT_{0R} \right)_{nn}   & = & \exp(ik_n d_{0R})              =  
\left( \bT_{R0} \right)_{nn}^*                           \; .
\label{diagonal2} 
\end{eqnarray}

A major simplification results in the class of decompositions where
there are no tubes on one side of $\Sigma_0$, as shown in 
Fig.~\ref{fig2}(b). Then $\bS$ reduces to 
\begin{equation}
\bS=\bR_{LL}= \bR_{LL}^0 + 
\bT_{L0} \left[ \one - \bR^R_{00} \bR^L_{00} \right]^{-1} 
\bR^R_{00} \bT_{0L} \;.
\label{scatback}
\end{equation}
This will be the matrix studied in this Sec.~\ref{sec:semiseparable}, 
since it already exhibits all the resonant structure of the general case. 
In the trivial case of a semi-infinite tube, Fig.~\ref{fig2}(c), 
again $\bR^L_{00}=0$ and $\bT_{0L}$ is given by (\ref{diagonal1}), 
whereas
\begin{equation}
\left( \bR^R_{00} \right)_{nn}   = -\exp(2i k_n d_R)   \; ,
\label{reflection}
\end{equation}
assuming Dirichlet boundary conditions at the end of the tube.

It is important to remember that the generalized $\bS$-matrices defined
in this section have infinite dimension. Only the blocks made up of
all the open channels will be unitary. For this reason,
the identification of the open channels
for reflection with a unitary $\bS$-matrix in (\ref{scatback}) is only valid 
because there are no open transmission channels. All projections of
a unitary matrix into a subspace define a dissipative matrix with
eigenvalues having moduli smaller than one, unless there is no
interaction with the subspace that has been projected away\cite{ozorio99}.
The addition of closed modes does not alter the
dissipative nature of the reflection matrices in the general case 
where transmission is present. 



\section{Resonance structure: multiple scattering}
\label{sec:resonance}


Let us assume that neither matrix $\bS_R$ or $\bS_L$ exhibits a sensitive
energy dependence. As previously discussed, such a situation can always
be reached in a sequence of decompositions. (Even if the scatterer has
a fractal structure, the wavelength settles the smallest scale of the
fractal that can be resolved, and thus the number of auxiliary sections
will be finite.) 
The sensitive energy dependence characteristic of a
resonant scatterer then emerges from the possibility that one of the
eigenvalues of $\bR^R_{00} \bR^L_{00}$ or $\bR^L_{00} \bR^R_{00}$ approaches
the value one. In both cases the {\em spectral determinant\/}
\begin{equation}
\det (\one- \bR^R_{00} \bR^L_{00})= 
\det (\one- \bR^L_{00} \bR^R_{00})       \to 0 \; .
\label{determinant}
\end{equation}

As we show in \cite{ozorio99}, any block with a smaller 
dimension than that
of a full unitary matrix will generally have eigenvalues with
moduli smaller than one. If the channels on the right are closed,
$\bR^R_{00}$ will be unitary, for the channels with real $k_0$, but the
product with $\bR^L_{00}$ will still be dissipative. So we never have
exactly zero in (\ref{determinant}) for scattering situations; indeed these 
singularities characterize bound states.
We may consider the bound states of closed systems as the limit of 
scattering systems in which the
channels both to the right and to the left are closed. The condition
(\ref{determinant}) is then precisely the well established condition 
for the existence of an eigenstate \cite{bogomolny92,rouvinez95,prosen95}.
Clearly, the order of the matrices $\bR^L_{00}$ and $\bR^R_{00}$ makes no
difference to the bound state theory, though it does affect the full
scattering problem. For the rest of this section this effect will be
unimportant, so we shall abreviate both products 
$\bR^R_{00} \bR^L_{00}$ and
$\bR^L_{00} \bR^R_{00}$ as simply $\bR_0$.

The fact that $\bR_0$ is a dissipative matrix allows us to expand 
\begin{equation}
\left( \one- \bR_0 \right)^{-1}=\sum_{\nu=0}^{\infty} \bR_0^\nu \; .
\label{geometric}
\end{equation}
We can thus interpret the formulae for each of the various scattering 
submatrices as the result of multiple reflections within the
resonant region, each one contributing a term to the total scattering
matrix. Note that (\ref{geometric}) only converges for a scattering system. 
For a bound system, this is only a formal equality. Therefore,
we can regularize the Bogomolny theory for a closed system by considering 
it as the limit of a scattering system.

The fact that $\bR_0$ is a well behaved operator with finite trace leads
to the Fredholm expansion
\begin{equation}
\left( \one- \bR_0 \right)^{-1}=\frac{1}{D}
\sum_{m=0}^{\infty} \sum_{k=0}^{m} D_k \bR_0^{m-k} \; ,
\label{fredholmexpansion}
\end{equation}
where the spectral determinant
\begin{equation}
D \equiv \det \left( \one- \bR_0 \right) = \sum_{m=0}^{\infty} D_m \; 
\label{fredholm}
\end{equation}
and
\begin{equation}
D_m = -\frac{1}{m} \sum_{k=1}^{m} D_{m-k} \mbox{Tr} \bR_0^k   \; .
\end{equation}
This is a similar expansion to that used by Georgeot and Prange
\cite{georgeot95}, but their theory is a reworking of the semiclassical 
theory, as compared to the exact result. 
In practice, one must work predominantly
with the open channels contemplated by the semiclassical theory (see
next section), but this finite matrix can be supplemented by as many
closed channels as necessary to obtain convergence.

The main point of the present theory is that we can define a 
{\em resonant envelope} for the energy dependence of all the matrix elements
of $\bS$ as the graph of $D^{-1}(E)$. Certainly, there is a complex
coupling among the different transverse states that varies smoothly with energy,
so that this graph is not strictly an envelope. Even so, it is only at
the peaks of $D^{-1}(E)$ that we will find sharp resonant amplitudes of
any element of the $\bS$ matrix. Not all the peaks of $D^{-1}(E)$ will
manifest themselves for a single matrix element, but it is not surprising
that collective properties of the $\bS$-matrix can be much more sensitive
to this function of the surface of section. 
Indeed, the {\em Wigner delay time},
\begin{equation}
\tau(E)= -\frac{i\hbar}{\Lambda} \frac{d}{dE} 
\log \det \bS_{\mbox{\scriptsize{open}}} \; ,
\end{equation}
where $\Lambda$ is the number of open channels in $\bS$, measures the 
globally resonant nature of the scatterer, {\em i.e.\/}, the 
peaks in $\tau(E)$ correspond to the energies of longest permanence in the
scattering region for some channel. Near such a peak, we may factor the
rapidly varying part of $\bS$ in (\ref{scatback}), or in the more general 
cases, as 
\begin{equation}
\bS_L \approx D^{-1}(E) \, \bsigma_L \; ,
\end{equation}
where $\bsigma_L$ has no peaks. It follows that near the resonant peaks
\begin{equation}
\tau(E) \approx -\frac{i \hbar}{\Lambda} \frac{d}{dE} 
                 \log ( D^{-\Lambda} \det \bsigma ) 
        \approx i\frac{d}{dE} 
                 \log D                         \; .
\end{equation}
Thus, we obtain the peaks of $\tau(E) \approx \tau_0(E)$, where we define
the {\em section time delay}
\begin{equation}
\tau_0(E) = -i \hbar \frac{d}{dE} \log \det (1-\bR_0)^{-1}         \; .
\label{sectiontime}
\end{equation}
We shall verify the validity of this approximate form of the time delay
for a numerical example in Sec.~\ref{sec:semiseparable}.


\section{Semiclassical approximations}
\label{sec:semiclassical}


The first step towards a semiclassical approximation of the preceeding
theory is the truncation of the generalized $\bS$-matrices, which are
restricted to the open channels. 
The reflections are then described by
square sub-matrices, whereas the blocks accounting for transmission
will be generally rectangular. 
Even though all the elements of (\ref{transref1}-\ref{transref4})
are then well defined, this approximation introduces spurious
discontinuities in their energy dependence as each new channel is
opened.  
The semiclassical theory can only avoid these by some form of
complex continuation of the classical motion. 
It should be noted that,
even so, the discontinuities in energy do reflect some of the
qualitative features of the full quantum theory, so that energy
integrals may be quite reasonable. 
In any case, we expect that
truncation will be a good approximation in an energy range with a
constant number of open channels, in the limit when this number is
large.

We now associate the transverse eigenstates $|n_J\}$ singled out 
by EBK-quantization (see e.g., \cite{ozorio89}) on a given section 
$\Sigma_J$ to the invariant tori for the classical Hamiltonian
$H_J(y_J,p_{y_J})$ defined by (\ref{hamiltonian}). 
In other words, we consider the canonical 
transformation $(y_J,p_{y_J})$ $\to$ $(I_J,\theta_J)$, such that
$H_J=H_J(I_J)$, obtained from the multivalued generating function
$F_\nu(I_J,y_J)$:
\begin{equation}
\frac{\partial F_\nu}{\partial y_J}=p_{y_J}  \; , ~~   
\frac{\partial F_\nu}{\partial I_J}=\theta_J \; 
\end{equation}
(where the index $\nu$ distinguishes the different branches). 
Then we approximate \cite{ozorio89}
\begin{equation}
\{y_J|n_J\} \approx \sum_\nu
\left| \frac{\partial^2 F_\nu}{\partial I_J \partial y_J} \right|
\exp \left[ i F_\nu (I_J,y_J)/\hbar \right] \; ,
\end{equation}
with 
\begin{equation}
I_J(n)=\hbar (n+\mu/4) \; ,
\end{equation}
where $\mu$ is the Maslov index for the torus. (The phase differences
on passing to a different branch are additive, having been 
included in $F_\nu$.) The number of open channels $\Lambda_J(E)$ is the 
largest integer $n$ satisfying the condition 
\begin{equation}
H_J(I_J(n)) \le E \; .
\label{restriction}
\end{equation}

\begin{figure}[ht]
\epsfxsize=8cm
\epsfbox[84 290 519 505]{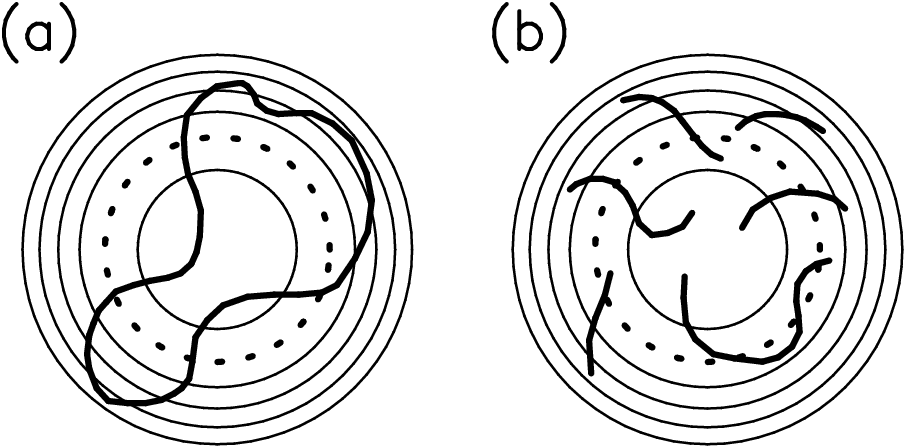}
\vspace{1cm}
\caption{ Single-lead scattering viewed on the surface of 
section. 
The circles represent channel tori selected by EBK quantization.
If scattering is nonresonant a torus (dotted) is mapped into a closed
curve [thick line in (a)]. In a resonant case (b), the mapping is
discontinuous and the torus returns to the section as a set of fragments. 
This set may be infinite and even fractal.}
\label{fig4}
\end{figure}

Let us first study scattering where all the points in $\Sigma_J$, subject 
to (\ref{restriction}), return to the same region. 
Semiclassically we have a reflection, $\bS=\bR$, corresponding to a 
conservative classical map. 
Each quantized torus determines a tube of
trajectories that re-intersect  $\Sigma_J$, i.e., it ``evolves''
while preserving its area as shown in Fig.~\ref{fig4}. 
The overlap of the evolved torus state with the basis of the quantized 
tori is just
\begin{equation}
\{n'_J|\bR|n_J\} \approx \sum_\nu
\left| \frac{\partial^2 F_\nu}{\partial I_J \partial I'_J} \right|
\exp \left[ i F_\nu (I_J,I_J')/\hbar \right] \; ,
\label{miller}
\end{equation}
where the canonical transformation is generated by
\begin{equation}
\frac{\partial F_\nu}{\partial I'_J} = -\theta'_J  \; , ~~   
\frac{\partial F_\nu}{\partial I_J}  =  \theta_J   \; .
\end{equation}
This general formalism for semiclassical maps was originally developed
by Miller \cite{miller74}. 
The branches, indexed by $\nu$, correspond to each transverse intersection 
of the new torus with the basis torus $I_J=\hbar (n+\mu/4)$. 
Both the new torus and the basis tori are closed curves (for nonresonant 
scattering) so there will be an even number of branches. 
Caustics, leading to spurious
singularities in (\ref{miller}), occur when the intersection of the
tori is non-transverse, i.e., where two intersections coalesce. 
In this region the matrix element should be expressed in terms of Airy
functions instead of (\ref{miller}) (see e.g., \cite{ozorio89}).
As we shall discuss below, this simple picture of a returning torus 
that overlaps with the basis of tori is fragmented into a fractal
mosaic for a very resonant system. However, we can now keep to a
simple
description for each nonresonant scatterer into which the section
decomposes the original resonant system.

It is important to note that we need not worry about the problem of
quantizing a compact phase space. Our torus basis is infinite, though
we are concerned with the projection within a classically invariant
region (which grows with $E$). The matrix elements among the open
channels in the semiclassical approximation depend only on the
classical dynamics within this region.  The discrete action variables
form a very privileged basis.  We cannot transform $\bR$ to the
(semiclassical) $|\theta_J\}$ basis without doing a Fourier sum over
elements $\{n'_J|\bR|n_J\}$ that lie outside the open region, not
defined semiclassically.  The same difficulty involves passing to the
$|y_J\}$ representation of $\bR$. However, the semiclassical evaluation
of unitary transformations relies entirely on the stationary phase
approximation. We can thus make ordinary semiclassical changes of bases
within the allowed region by assuming that there are no stationary
phase points outside it.

\begin{figure}[ht]
\epsfxsize=8cm
\epsfbox[84 266 519 527]{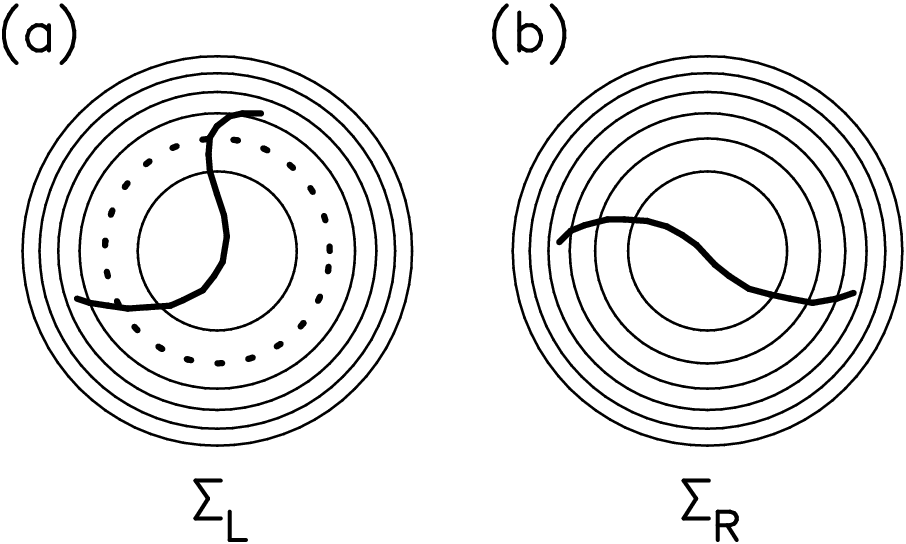}
\vspace{1cm}
\caption{In multi-lead scattering the full surface of section is the
union of the sections of all leads (two in this example). Shown is a 
torus in the left section [dotted, (a)] which is divided into a 
reflected part and a transmitted part [thick lines in (a) and 
(b), respectively].}
\label{fig5}
\end{figure}

For scattering where the open channels are not restricted to a single
lead, we must consider the conservative classical map defined on the 
union of several surfaces of section, as shown in Fig.~\ref{fig5}. 
Again, the classical motion takes place on the region (\ref{restriction}) 
for each section, so that the allowed region will vary from section to 
section. 
The basis of eigenstates for each section corresponds to a set of quantized
invariant tori of that particular section.

To determine the semiclassical approximation to the $\bS$-matrix when
there is more than one lead, we again
consider the tube of trajectories defined by one of the quantized tori in
one of the sections. But now this tube will generally split, reintersecting the 
various sections along open segments. Semiclassically, there will be
nonzero matrix elements with all the quantized tori with which the
segments intersect. The elements of the reflection block are still given
by (\ref{miller}), whereas the transmission elements from the $J$-lead to the
$K$-lead depend of the generating functions $F_\nu(I_J,I_K)$ for the
canonical transformation $(I_J,\theta_J) \to (I_J,\theta_J)$ in the form
\begin{equation}
\{n'_K|\bT|n_J\} \approx \sum_\nu
\left| \frac{\partial^2 F_\nu}{\partial I_J \partial I_K'} \right|^{1/2}
\exp \left[ i F_\nu (I_J,I_K')/\hbar \right] \; .
\label{tmiller}
\end{equation}

By adopting fixed action-angle variables along $\Sigma_J$, with the normal
coordinate $x_J$, the action for orbits returning to the same lead,
\begin{equation}
F_\nu (I_J,I_J')= \int_{I_J}^{I_J'} \theta_J dI_J + 
                  \oint p_{x_J}dx_J                      \; ,
\end{equation}
except for Maslov indices, is evaluated along the classical orbit
joining the torus $I_J$ to $I_J'$. This can also be obtained by
taking 
\begin{equation}
F_\nu (y_J,y_J')= \int  p_{y_J}dy_J + 
                  \oint p_{x_J}dx_J                      
\end{equation}
and then evaluating the change of basis by stationary phase. 
To evaluate 
the transmission actions, we first note that the difference between 
(\ref{miller}) and (\ref{tmiller}) is only that $I_J$ and $I_K'$ 
belong to different torus bases, rather than they belong to 
different sections. Defining $F^0$ as the generating function for
the canonical transformation corresponding to this change of
basis we obtain  
\begin{equation}
F_\nu (I_J,I_K')= 
F_\nu^0 (I_J,I_K') + \int_{I_K(I_J)}^{I_K'} \theta_K dI_K + 
                  \int_{\Sigma_J}^{\Sigma_K} p_{x}dx            \; ,
\end{equation}
where
\begin{equation}
\frac{\partial F_\nu^0}{\partial I_K'} = -\theta'_K    
\end{equation}
describes the original torus with action variable $I_J$ in the
$(I_K,\theta_K)$ coordinates; $I_K(I_J)$ is the action variable
of the orbit on this torus that will arrive at $I_K'$ and the
variable $x$ in the last integral is assumed normal to both 
sections.

So far, this review of semiclassical scattering theory applies to
an arbitrary scatterer. The problem arises, for a very resonant 
scatterer that the classical map fragments into arbitrarily small
regions which cannot be quantized when their area is smaller that
Planck's constant \cite{ozorio99}. However, we have seen 
that it is always possible to decompose a resonant scatterer by
appropriate surfaces of section. It is then possible to adopt
semiclassical approximations for each of the components and then
to evaluate the blocks of the full $\bS$ matrix given by 
(\ref{transref1}-\ref{transref4})
in the stationary phase approximation.

For a typical resonant scatterer, cut by a ``good'' section, $\Sigma_0$,
the classically allowed area in $\Sigma_0$ will be considerably
larger than the union of the $R$-leads and of the $L$-leads
projected onto $\Sigma_0$. The classical 
map corresponding to the various $\bT$ blocks of the component scatterers
will be a simple injection of orbits into $\Sigma_0$, or its time reverse.
If we decouple this part of the $\bS$ matrix, there results the view
of semiclassical scattering as essentially that of the $\Sigma_0$
surface onto itself with projections onto entrance and exit regions,
as previously proposed \cite{ozorio99}. 
The semiclassical neglect of any backscattering
in the $L0$ or $R0$ scattering depends on the smoothness of the Hamiltonian.
It is certainly inadequate for a discontinuity in the Hamiltonian within the
range of the longitudinal wavelength. This case will be treated in 
Sec.~\ref{sec:discontinuities}.

If we apply the Fredholm theory in Sec.~\ref{sec:decomposition} to 
$[1-\bR^R_{00} \bR^L_{00}]^{-1}$
and to $[1-\bR^L_{00} \bR^R_{00}]^{-1}$ and then use the semiclassical
expressions for each of the reflection matrices, we succeed in rederiving 
the scattering theory of Georgeot and Prange \cite{georgeot95} from first 
principles.
This is especially valuable, because it was originally obtained by the
rearrangement of the semiclassical contributions to the scattering 
matrix, which is not valid in the resonant context. The spectral determinant
(\ref{fredholm}) will be described by the classical periodic orbits 
that cross the 
surface $\Sigma_0$ rearranged into composite orbits or pseudo-orbits.
As we have seen, the spectral determinant is never zero, because the
$\bR$ matrices are dissipative. 
No matter how long lived the typical
classical orbits may be, as counted by the number of traversals of
$\Sigma_0$, there will be a cuttoff for the period of the periodic
orbits included in the Fredholm theory. This is limited to the dimension
of the $\bR$ matrices.

\begin{figure}[ht]
\epsfysize=5cm
\epsfbox[102 248 510 551]{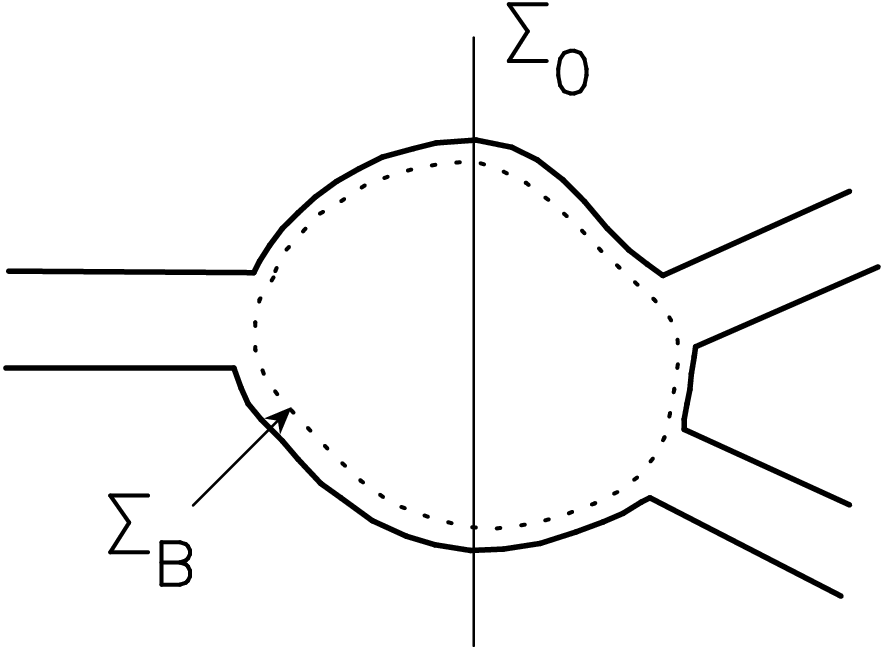}
\vspace{1pc}
\caption{A billiard connected to leads. Plot are a ``standard'' section
$\Sigma_0$ and the Birkhoff section $\Sigma_B$. Thick lines represent
the contour of the open billiard.}
\label{fig6}
\end{figure}

To conclude this section, we note that in the important case of systems
with hard walls, i.e., open billiards connected by leads as in 
Fig.~\ref{fig6},
the quantized tori for any section, central to the foregoing theory,
will be defined as the level curves of $p_y^2$. 
The phase change between the two branches of $\{y|n\}$ will depend upon the 
choice of Dirichlet or Neumann boundary conditions. It may well be 
advantageous to use the perimeter of the closed billiard 
($\Sigma_B$ in Fig.~\ref{fig6}),
corresponding to the classical Birkhoff map (or bounce map) instead of
a section such as $\Sigma_0$. This was the point of view adopted in 
\cite{ozorio99}.
In this case, all scattering will be considered as a reflection across
$\Sigma_B$. However, in this case, it is essential to take into account
the discontinuity on entering the billiard, which is the subject of the 
next section.


\section{Discontinuities}
\label{sec:discontinuities}

\begin{figure}[ht]
\epsfysize=5cm
\epsfbox[80 229 533 572]{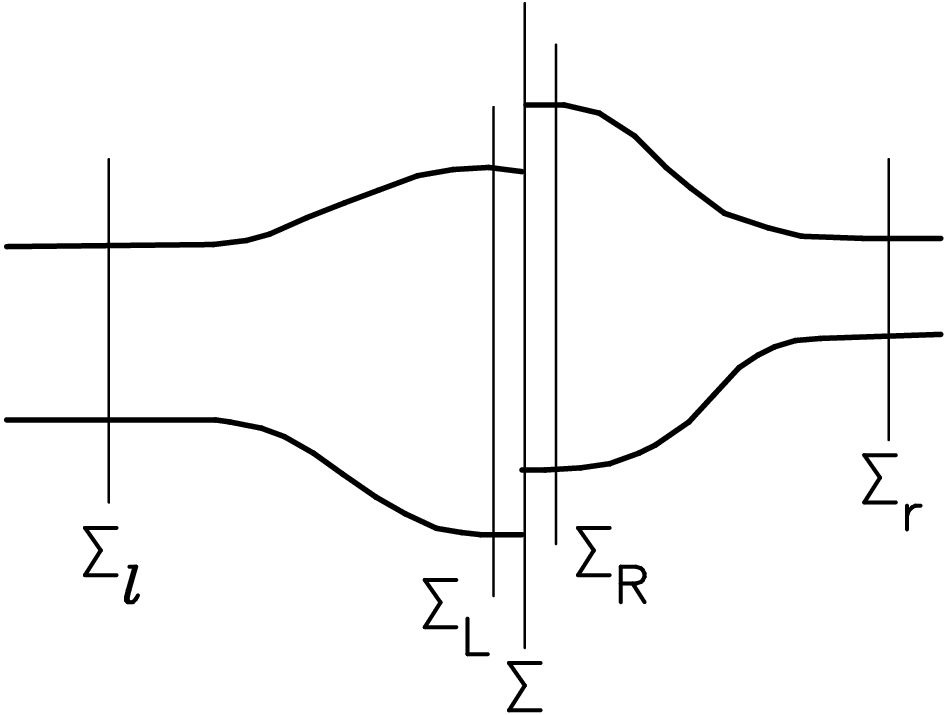}
\vspace{1pc}
\caption{A scattering system discontinuous across a surface $\Sigma$. 
The sections $\Sigma_L$ and $\Sigma_R$ allow for isolating the problem
of the discontinuity in the scattering from $\Sigma_l$ to $\Sigma_r$.}
\label{fig7}
\end{figure}

If the Hamiltonian is discontinuous across a given plane $\Sigma$,
we should construct two sections $\Sigma_L$ and $\Sigma_R$ immediately
on either side of $\Sigma$, as shown in Fig.~\ref{fig7}.
This is an important example of a scattering system which is
clarified by the use of extra surfaces of section.
In the following theory we only address the local problem of scattering
from $\Sigma_L$ to $\Sigma_R$. With respect to Fig.~\ref{fig7}, we
may consider the $l \to R$ scattering as decomposed by $\Sigma_L$, or 
that $\Sigma_R$ decomposes the $L \to r$ scattering. Thus the full
$l \to r$ scatterer will be decomposed in two stages. In the simple
semiseparable example discussed in Sec.~\ref{sec:semiseparable}, 
both the $l \leftrightarrow L$ and $R \leftrightarrow r$ 
propagators are trivial.

To solve the scattering problem across the discontinuity between the 
sections $\Sigma_L$ and $\Sigma_R$ involves the same smoothness 
conditions as (\ref{matchk1}) and (\ref{matchk2}), except that now the wave 
vectors $\bk_R \ne \bk_L$ and we have different basis states $|n_R\}$ and
$|n_L\}$ to match on either side. Recalling the definiton of the
operators $\bk_J^\alpha$ in (\ref{kalpha}), we have
\begin{eqnarray}
\bk_L^{-1/2} \left[ | \psi_L \}^+  + | \psi_L \}^- \right] & = &
\bk_R^{-1/2} \left[ | \psi_R \}^+  + | \psi_R \}^- \right] \; , \\ 
i \bk_L^{1/2} \left[  | \psi_L \}^+  - | \psi_L \}^- \right] & = &
i \bk_R^{1/2} \left[ -| \psi_R \}^+  + | \psi_R \}^- \right] \; .
\end{eqnarray}
Decomposing the $\bS$ matrix into reflexion and transmission 
blocks (\ref{matdef}), then yields
\begin{eqnarray}
\bk_L^{-1/2} \left[ (\one + \bR_{LL}) |\psi_L \}^- +
                       \bT_{LR} |\psi_R \}^- 
           \right] & = &
\bk_R^{-1/2} \left[ (\one + \bR_{RR}) |\psi_R \}^- +
                       \bT_{RL} |\psi_L \}^- 
           \right] \; , 
\label{discon1} \\
\bk_L^{1/2}  \left[ (\one - \bR_{LL}) |\psi_L \}^- -
                       \bT_{LR} |\psi_R \}^- 
           \right] & = &
\bk_R^{1/2}  \left[(-\one + \bR_{RR}) |\psi_R\}^- +
                       \bT_{RL} |\psi_L\}^- 
           \right] \; .
\label{discon2}
\end{eqnarray}

The fact that these equations are valid for any  
$|\psi_R\}^-$ or $|\psi_L\}^-$ allows us to equate separately 
the operators acting on either of these functions. However,
to derive matrix equations, we must transform between the 
natural bases on either side. Thus, defining the orthogonal
matrix
\begin{equation}
\bU_{n'n}= \{ n'_L | n_R \} = \int dy \{ n'_L|y \} \{ y|n_R \} \; ,
\label{overlap}
\end{equation}
we obtain
\begin{eqnarray}
\bk_L^{-1/2} (\one + \bR_{LL}) & = & \bU \bk_R^{-1/2} \bT_{RL} \; ,
\label{matchdisc1} \\
\bk_L^{ 1/2} (\one - \bR_{LL}) & = & \bU \bk_R^{ 1/2} \bT_{RL} \; ;
\label{matchdisc2}
\end{eqnarray}
where the matrix $\bR_{LL}$ is in the $L$-representation, whereas 
$(\bT_{RL})_{nn'}=\{ n_R |\bT_{RL} | n_L' \} $. 
Alternatively, we obtain from (\ref{discon2}) and the transpose of
(\ref{overlap}) that
\begin{eqnarray}
\bk_R^{-1/2} (\one + \bR_{RR}) & = & \bU^T \bk_L^{-1/2} \bT_{LR} \; , 
\label{matchdisc3} \\
\bk_R^{ 1/2} (\one - \bR_{RR}) & = & \bU^T \bk_L^{ 1/2} \bT_{LR} \; ,
\label{matchdisc4}
\end{eqnarray}
where, again, the indices $L$,$R$ specify the bases. Notice that
the set of equations (\ref{matchdisc3},\ref{matchdisc4}) (and its 
solution) is related to the set (\ref{matchdisc1},\ref{matchdisc2}) 
by interchanging $L,\bU \leftrightarrow R,\bU^T$(time reversal). 
So we restrict ourselves to (\ref{matchdisc3},\ref{matchdisc4}), 
which can be solved for 
\begin{eqnarray}
\bT_{LR} & = & 2 \left( \bk_R^{ 1/2} \bU^T \bk_L^{-1/2} +  
                      \bk_R^{-1/2} \bU^T \bk_L^{ 1/2} 
               \right)^{-1}                         \; , 
\label{discT}                                            \\
\bR_{RR} & = & {\textstyle \frac{1}{2}}
               \left( \bk_R^{ 1/2} \bU^T \bk_L^{-1/2} -  
                      \bk_R^{-1/2} \bU^T \bk_L^{ 1/2}
               \right) \bT_{LR}                       \; ;
\end{eqnarray}
with $\bT_{RL}$ and $\bR_{LL}$ given by time-reversal as we previously
noted:
\begin{eqnarray}
\bT_{RL} & = & 2 \left( \bk_L^{ 1/2} \bU \bk_R^{-1/2} +  
                      \bk_L^{-1/2} \bU \bk_R^{ 1/2} 
               \right)^{-1}                         \; , \\
\bR_{LL} & = & {\textstyle \frac{1}{2}}
               \left( \bk_L^{ 1/2} \bU \bk_R^{-1/2} -  
                      \bk_L^{-1/2} \bU \bk_R^{ 1/2}
               \right) \bT_{RL}                       \; .
\end{eqnarray}
It is straightforward to verify that the $\bS$ matrix
for the discontinuity is symmetric. The verification of unitarity
is also simple, but lengthy\cite{weidenmuller64}. 

The matrix inversions in the preceding formulae will be calculated
by truncating the bases on the right and on the left of the
discontinuity. A possible choice is to limit $\bU$ to the open channels
on either side, but their numbers may not be the same. A better
alternative is to notice that the matrix elements $\bU_{n'n}$ decay 
exponentially when there is no intersection between the quantized tori
corresponding to $|n_L\}$ and to $|n'_R\}$. Therefore, the open channels
should be supplemented with states which intersect with them on either 
side of the discontinuity.

\begin{figure}[ht] 
\epsfysize=5cm
\epsfbox[71 290 500 527]{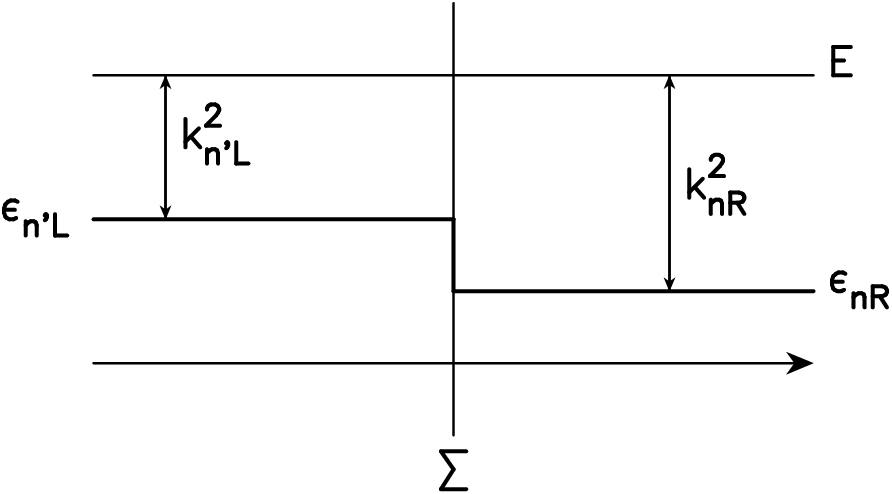}
\vspace{1pc}
\caption{The ``equivalent'' one dimensional scattering problem for the
discontinuity. A particle with energy $E$ propagates from the right
channel $nR$ to the left channel $n'L$ through the discontinuity 
$\Sigma$. The jump in the longitudinal momentum is $k_{n'L}-k_{nR}$,
where $\hbar^2 k_{n'L}^2/2m=E-\epsilon_{n'L}$ and $\hbar^2 k_{nR}^2/2m=E-\epsilon_{nR}$.}
\label{fig8}
\end{figure}

It is interesting that the preceeding argument relies on a semiclassical
criterion of torus overlap to calculate $\bU_{n'n}=\{n'_L|n_R\}$, even
though we use this to calculate classically forbidden reflections.
We shall use a similar semiclassical approach to estimate the channels
exhibiting resonances in Sec.~\ref{sec:semiseparable}.
This semiclassical inspiration will now be pushed further to suggest
an approximation that avoids the matrix inversions in $\bS$. To this end,
we notice that the semiclassical evaluation of the nondecaying $\bU_{n'n}$
depends on the points of intersection for the corresponding tori
(such as shown in Fig.~\ref{fig10}). We could then view each intersection as
the projection of an orbit transmitted or reflected at the discontinuity.
Thus each orbit contributing to a given matrix element faces the same
discontinuity in the longitudinal 
Hamiltonian, as depicted in Fig.~\ref{fig8}.
What is the probability amplitude for a wave $|n_R\}$ incident on the right
to propagate to the left channel $|n'_L\}$?
Separating out the longitudinal motion, we have  
\begin{eqnarray}
|x,y_R \rangle & = & \left[ k_{nR}^{-1/2} \exp(-ik_{nR}x) + 
                     (\br_{RR})_{n'n} k_{nR}^{-1/2} \exp(ik_{nR}x) 
                    \right]
                     \{y|n_R \} \; , x \ge 0 \; ,  \nonumber \\
|x,y_L \rangle & = & \left[ (\bt_{LR})_{n'n} k_{n'L}^{-1/2} \exp(-ik_{n'L}x) 
                    \right]
                     \{y|n'_L\} \; , x \le 0 \; .
\label{unimatch}
\end{eqnarray}
Here, $\bt_{LR}$ and $\br_{RR}$ are the ``longitudinal'' transmission and
reflection matrices. By requiring that both waves match at the discontinuity
at $x=0$ we obtain:
\begin{eqnarray}
 (\bt_{LR})_{n'n} & = &  2  \left( \sqrt{\frac{k_{nR }}{k_{n'L}}} + 
                                   \sqrt{\frac{k_{n'L}}{k_{nR }}} 
                            \right)^{-1}  \; ,                        \\
 (\br_{RR})_{n'n} & = &     \left( \sqrt{\frac{k_{nR }}{k_{n'L}}} - 
                                   \sqrt{\frac{k_{n'L}}{k_{nR }}} 
                            \right)
                            \left( \sqrt{\frac{k_{nR }}{k_{n'L}}} + 
                                   \sqrt{\frac{k_{n'L}}{k_{nR }}} 
                            \right)^{-1}\; ,
\label{uniscat}
\end{eqnarray}
which are the well known results for the one-dimensional step
potential.  In the case of a two-dimensional discontinuity we still
have to take into account the coupling $U_{n'n}$ between transverse
modes at each side of the discontinuity.  The expression for the
approximate transmission matrix $\widetilde{\bT}$ is thus
\begin{equation}
(\widetilde{\bT}_{LR})_{n'n} = (\bt_{LR})_{n'n} \bU_{n'n} \equiv 
                               \left[ \bt_{LR} \bU \right]_{n'n} \; .
\label{apptrans}
\end{equation}
Similarly $(\br_{RR})_{n'n} \bU_{n'n}$ gives the probability amplitude for
{\em not being transmitted} to channel $n'$. As this probability is distributed 
over the $L$ basis we must switch back to the $R$ representation. Our approximate
result for the reflection matrix reads
\begin{equation}
\widetilde{\bR}_{RR} = \bU^T \left[ \br_{RR} \bU \right] \; .
\end{equation}
The corresponding approximation for the transmission and reflection matrices 
for waves inciding from the left are obtained by the time reversal operation:
\begin{eqnarray}
\widetilde{\bT}_{RL} & = &      \left[ \bt_{RL} \bU^T \right]  \; , \\
\widetilde{\bR}_{LL} & = &  \bU \left[ \br_{LL} \bU^T \right]  \; .
\end{eqnarray}
\noindent
We can clarify the nature of this two dimensional version of a ``sudden
approximation'' (\ref{apptrans}) by noting that 
\begin{equation}
\bk_L^{-1/2} \widetilde{\bT} \bk_R^{ 1/2} \bU^T + 
\bk_L^{ 1/2} \widetilde{\bT} \bk_R^{-1/2} \bU^T = 2 \; 
\end{equation}
as compared to (\ref{discT}), so that $\widetilde{\bT}$ amounts to a
reordering of the operators that define $\bT$. This allows the
simplification of not having to invert the matrices in the exact
formula (\ref{discT}).


\section{A semiseparable numerical example}
\label{sec:semiseparable}

\begin{figure}[ht]
\epsfysize=8cm
\epsfbox[80 204 519 580]{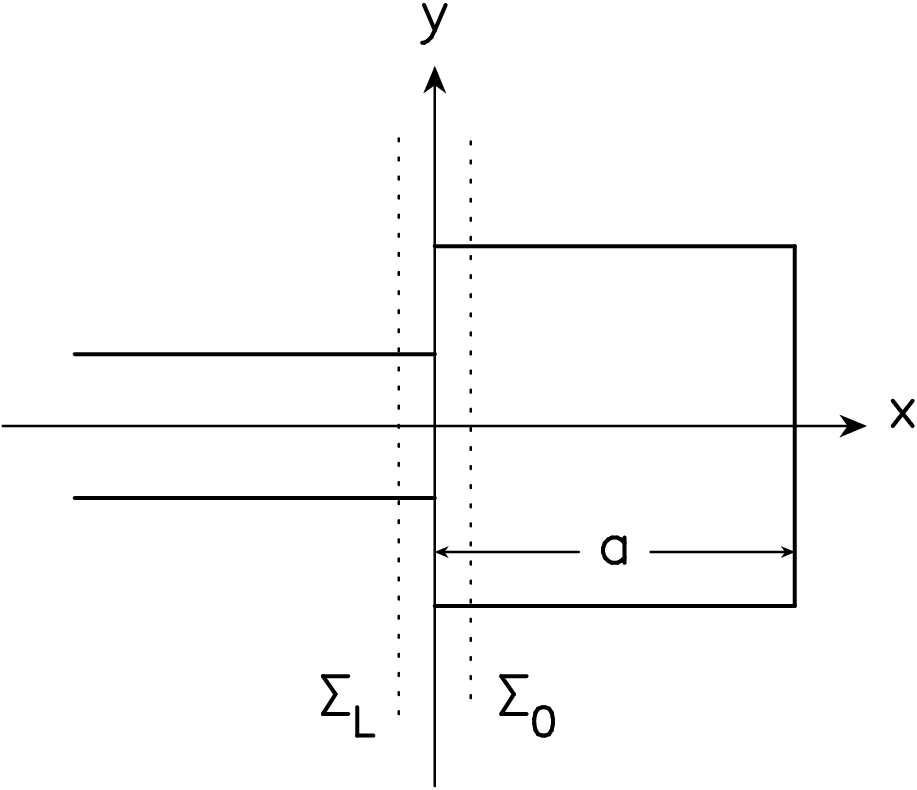}
\vspace{1pc}
\caption{The semiseparable system consists of two waveguides: a semi-infinite
one ($x<0$) and a finite one ($x>0$). Dirichlet conditions are assumed
at $x=a$. Also shown are the auxiliary sections $\Sigma_L$ and $\Sigma_0$.}
\label{fig9}
\end{figure}

We now study a simple though nontrivial example of the decomposition of
a resonant scatterer. This is chosen as the semiseparable system sketched 
in Fig.\ref{fig9}, {\em i.e.\/}, the potential $V(x,y)$ is separable for
$x>0$ and for $x<0$, but discontinuous at $x=0$:
\begin{equation}
V(x,y) =
\left\{ \matrix{ \half \omega_L^2 y^2 &  x<0     \cr
                 \half \omega_R^2 y^2 &  0<x<a   \cr
		 \infty               &  x=a         } 
\right. \; .~
\label{potential}
\end{equation}
\noindent
We choose $\omega_L > \omega_R$, so that the classical motion
is broader on the right than on the left: the outline in 
Fig.~\ref{fig9} represents an equipotential curve. 
(The main difference with respect to the system studied by 
 Prosen\cite{prosen95} is that it is open in the left.)
 Computations were carried out with the values $\omega_L$=1,
 $\omega_R$=0.71423, $a$=30, $\hbar$=1, and mass $m$=1 (in
 appropriate units).
 
 Separability on the left allows us to bring the entrance section
 $\Sigma_L$ to $x=0^-$. The position of the Poincar\'e section is
 also arbitrary, because of separability, so we bring
 $\Sigma_0$ to $x=0^+$. There being a single open lead, the
 decomposition of the scattering matrix is given by (\ref{scatback}),
 where the reflection matrix on the right of $\Sigma_0$ is diagonal,
 with elements
\begin{equation}
 \left\{n_R | \bR^R_{00} | n_R \right\} = -\exp( 2ik_{Rn} a) \; .
\end{equation}

 The transmission matrices $\bT_{L0}$ and $\bT_{0L}$ are just those
 for the passage through a discontinuity of the potential discussed
 in the last section. This permits us to evaluate the accuracy of
 the approximation proposed there within the resonant theory. The 
 reflection $\bR^R_{00}$ may also be calculated in this approximation.
 In spite of its motivation in terms of classical orbits, one should
 notice that this ``sudden approximation'' mixes the motion that fails
 to escape before it is propagated diagonally by $\bR^R_{00}$. Thus,
 the full Poincar\'e reflection 
 matrix $\bR^R_{00} \bR^L_{00}$ is not diagonal, in contrast to the 
 classical motion which is integrable until it escapes.
 
 The $\bS$ matrix was calculated in an energy range corresponding to
 15 open channels, but excluding an interval of 0.2 near the thresholds
 of the 15th and 16th channels. Evidently each channel threshold 
 is given by 
\begin{equation}
 E_n = \hbar \omega_L (n-\half) \; .
\end{equation}
 The number of transverse states on the right corresponding to open
 channels grows from 21 to 22 in the energy window, with our choice 
 of the frequency $\omega_R$.
 
\begin{figure}[ht]
\epsfysize=8cm
\hspace{2cm}
\epsfbox[133 272 520 537]{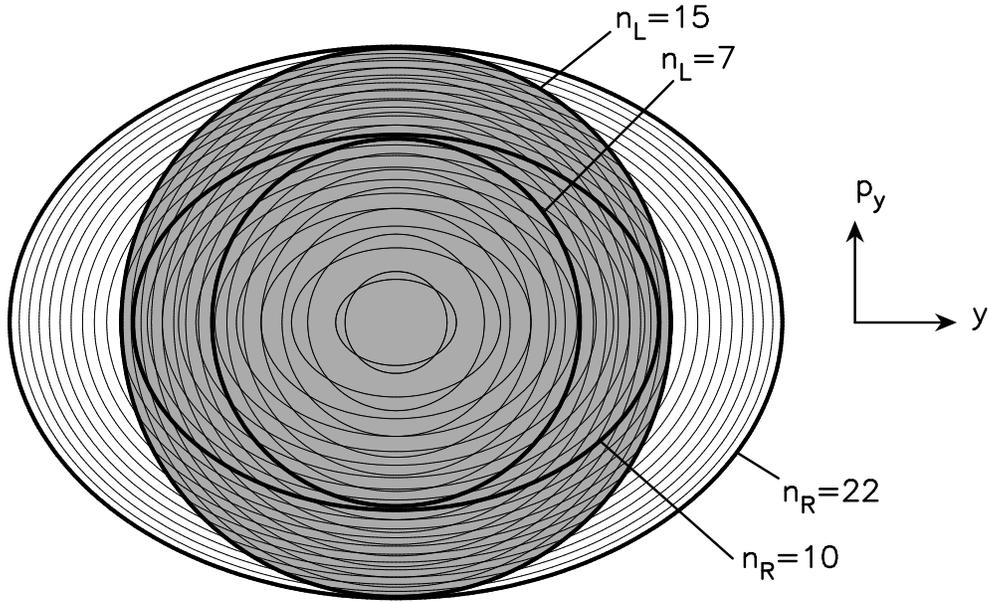}
\vspace{1pc}
\caption{Quantized tori on each side of the semiseparable system.
Circles correspond to channel modes and ellipses to cavity modes.
The outermost circle corresponds to the last open channel.}
\label{fig10}
\end{figure}

 We can predict which entrance channels may exhibit resonance behaviour 
 by reverting to the qualitative semiclassical picture of 
 Sec.~\ref{sec:semiclassical}, even though the discontinuity prevents
 an accurate semiclassical calculation. The modes in the open lead
 correspond to the concentric circles of in Fig.~\ref{fig10} with
 area $2\pi (n_L-\half)$, whereas the states in the cavity correspond
 to the ellipses with area $2\pi (n_R-\half)$ and semiaxes with the
 ratio $\omega_R$. 
 Semiclassically, an incoming mode only excites
 the inner states corresponding to ellipses that intersects its circle.
 If these ellipses, in their turn, only intersect circles with
 $n_L \le 15$, there should be straight nonresonant backscattering 
 for this mode. It is easy to see that, by this criterion, it is 
 only for $n_L>15\omega_R^2 \approx 7$ that the resonant structures
 should arise.
 
\begin{figure}[htbp]
\epsfysize=12cm
\epsfbox[45 216 511 540]{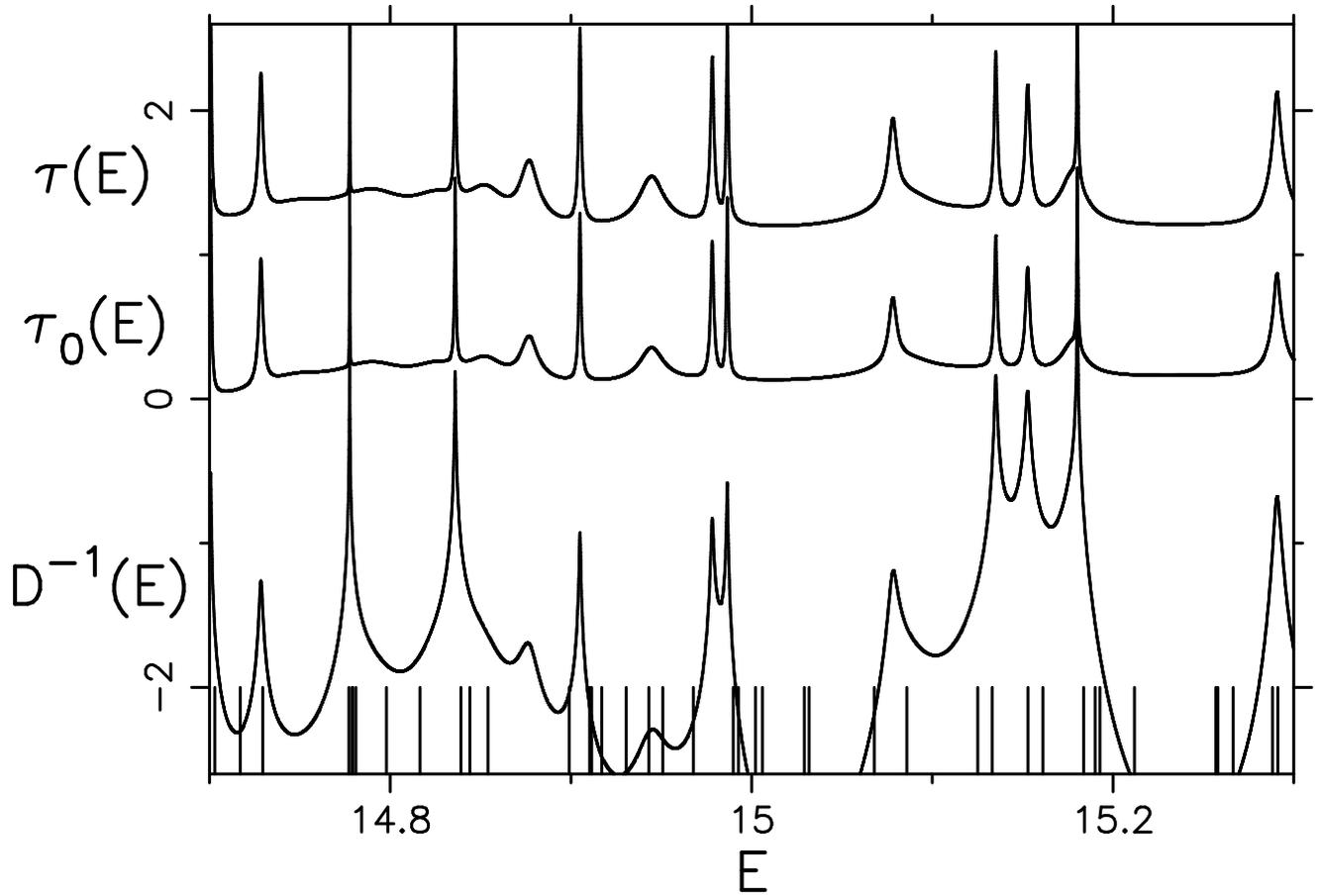}
\vspace{1cm}
\caption{Log$_{10}$-linear plot of the exact time delay $\tau(E)$, 
the section time delay $\tau_0(E)$, 
and the spectral determinant $D(E)$. 
The spikes at the bottom represent the energy levels of the
bound system obtained by closing the cavity with a hard wall at
$x=0$.}
\label{fig11}
\end{figure}

In Fig.\ref{fig11} we compose $D=\det[1-\bR^R_{00}\bR^L_{00}]$ and the
section time delay with Wigner's time delay. It is impressive how the
overall structure of this global imprint of the resonances is
re-composed from both nonresonant matrices $\bR^R_{00}$ and
$\bR^L_{00}$.
It is important to note that the peaks in the time delay are only
poorly correlated to the bound states of the separable system obtained
by closing the lead at $x=0$, also shown in Fig.~\ref{fig10}. [A
comment about this plot is in order. Recalling that in
(\ref{sectiontime}) we had discarded the slowly varying part of the
time delay, we shifted $\tau_0(E)$ by a fixed amount so that both
average times $\langle \tau_0 \rangle $ and $\langle \tau \rangle $ 
coincide.  Then logarithms were
taken and, finally, the curves corresponding to $\tau_0(E)$ and $D(E)$
were shifted downwards to make comparisons easier.]

\begin{figure}[ht]
\epsfysize=14cm
\epsfbox[37 70 523 468]{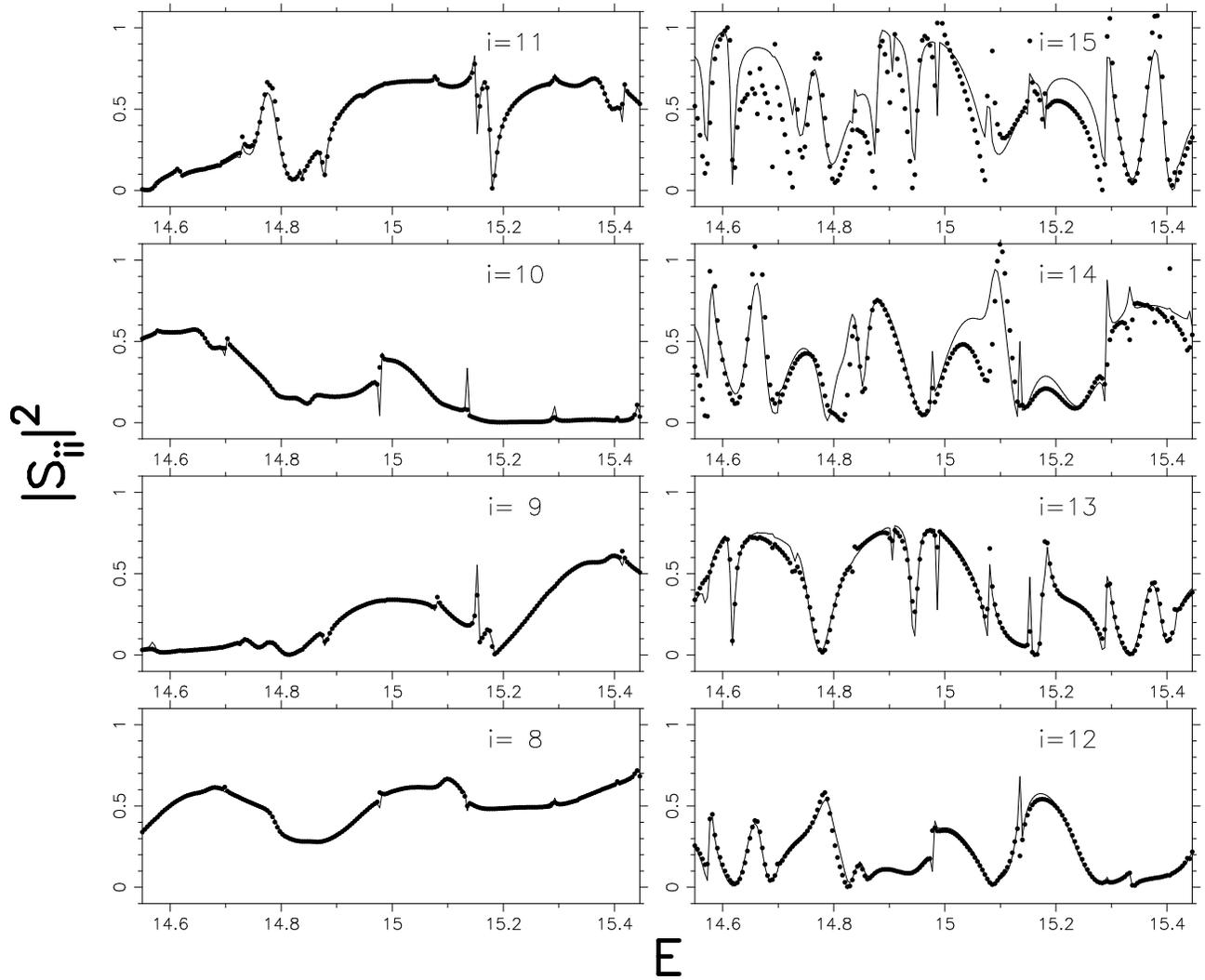}
\vspace{1cm}
\caption{Elastic cross sections $|\bS_{ii}|^2$ as a function of energy.
Full lines correspond to the exact calculations. Dots represent the
results of the ``sudden approximation'' (see text).}
\label{fig12}
\end{figure}

In Fig.\ref{fig12} we show the resonance structure of individual 
$\bS$ matrix elements. 
Evidently, each one of these does not exhibit all the resonance peaks
in the time delay, but the determinant $D$ does limit the positions 
allowed for these peaks in all cases. We find that the 
positions of the resonances are reasonably obtained, except for very 
fine structures, but the amplitudes of the individual matrix elements
are not predicted by the spectral determinant.
In these figures we also compare
the sudden approximation with the exact calculation. This approximation 
works remarkably well for both nonresonant and resonant channels up
to $i$=13.


\section{Concluding remarks}
\label{sec:concluding}


The spiky energy dependence of the $\bS$ matrix characteristic of
resonant scattering in quantum mechanics corresponds to complex orbital
structure of the classical limit.  In both theories these complications
can be explained in terms of the multiple iterations of a relatively
simple mapping defined on an appropriate surface of section. The
quantum theory for such an open system is based on the pioneering
developments of Bogomolny, Prosen and several papers by Smilansky and
coworkers on bound systems, but here we have the advantage that the map
is dissipative. Thus, our formula (\ref{fredholmexpansion}) converges
and we could, in principle, regularize the section theory for bound
systems by considering them as the limit of a family of open systems.

By focusing the scattering problem on a section map, there emerges the
central role of the spectral determinant.  As the openings of the
scatterer are closed, the complex zeroes of $D(E)$ converge onto the
real eigenergies of the resulting bound system. We have shown that even
for the open system the full profile of this energy function is found
to portray the qualitative features of the scattering time-delay. Our
numerical example reveals that, even for individual elements of the
$\bS$ matrix, the possible resonance peaks are restricted to those of
the spectral determinant.

The picture of the on-the-shell scattering as relating sections on
various leads, which may be decomposed by splicing the scatterer and
adding more leads, has a clear semiclassical interpretation. For a
scattering system with two degrees of freedom, there is only one
freedon left transverse to the leads. The corresponding classical
Hamiltonian is therefore integrable and this allows us to associate an
invariant torus (closed curve) of the asymptotic Hamiltonian to each
scattering channel. The classical propagation of each torus through the
scatterer generally breaks up these tori and the elements of the $\bS$
matrix result from the intersection of their fragments with the exit
channels. This theory goes back to Miller, but its resummation based on
a section was achieved by Georgeot and Prange. The advantage our new
derivation of this theory is that it proceeds from first principles,
rather than as a rearrangement of Miller's orbit sum.  Only in this way
can we show that the approximate semiclassical spectral determinant is
based on a dissipative rather than a unitary matrix.


\acknowledgements

The authors have benefited from discussions with C. H. Lewenkopf and
M. Saraceno. This work was supported by Brazilian agencies 
PRONEX, CNPq and FAPERJ.


%
%

\newpage

\newpage

\newpage

\newpage

\newpage

\newpage

\end{document}